\documentclass[12pt,preprint]{aastex}

\shorttitle{Surface gravity for a 
planetary member in the \sigori~ cluster}
\shortauthors{Mart\'\i n \& Zapatero Osorio}

\newcommand{\mjup}{\,$M_{\rm Jup}$}
\newcommand{\msol}{\,$M_{\odot}$}
\newcommand{\sigori}{$\sigma$\,Orionis}

\begin{document}

\title{Spectroscopic estimate of surface gravity for a 
planetary member in the \sigori~ cluster}

\author{E. L. Mart\'{\i}n\altaffilmark{1}}  
\affil{Institute of Astronomy. University of Hawaii at 
Manoa. 2680 Woodlawn Drive, Honolulu, HI 96822, USA}

\author{M. R. Zapatero Osorio\altaffilmark{2} }
\affil{LAEFF-INTA, Villafranca del Castillo, PO Box 50727, 
E-28080 Madrid, Spain}


\begin{abstract}

We present intermediate-resolution (R=1500) near-infrared spectroscopy 
from 1.17~$\mu$m to 1.37~$\mu$m of the spectral type T 
planetary candidate member in the 
\sigori~ cluster S\,Ori\,70 reported by Zapatero Osorio et al. (2002). 
The new data have been obtained with NIRSPEC at the Keck~II telescope. 
The best fit of our mid-resolution spectrum of S\,Ori\,70 with theoretical spectra gives log~g=3.5$\pm$0.5~cm\,s$^{-2}$ and T$_{\rm eff}$=1,100$^{+200}_{-100}$~K. 
The low gravity of this object derived from spectral synthesis 
supports its youth and membership 
to the young  \sigori~ cluster. Using 
evolutionary models for an age of 3~Myr, we obtain 
a mass of 3~\mjup ~and a radius of 0.16~$R_{\odot}$, independent of the 
distance to the object.  
Our analysis confirms that S\,Ori\,70 is the lowest mass cluster planet so 
far identified in the galaxy.   

\end{abstract}


\section{Introduction}

Objects with masses below the substellar mass limit ($M\,<\,0.072$\msol) 
fail to stabilize on the main-sequence and cool down to very low temperatures. 
Their observational study has started only recently, but it is progressing 
vigorously. A focus point in the research about substellar mass objects (SMOs) 
is the minimum 
mass that free-floating objects in young clusters and associations can have.
Particularly,  
it has been suggested that the faintest and coolest objects discovered through 
deep imaging in Orion have masses    
below the deuterium-burning limit at 
0.013\msol~or $\sim$ 14\mjup, where 1 \msol~= 1047 \mjup 
(Lucas \& Roche 2000; Zapatero Osorio et al. 2000; Lucas et al. 2001). 
Such mass limit has been used in the literature to distinguish between brown 
dwarfs (BDs) and planets (Burrows et al. 1997). 
Another mass limit that has been 
discussed lies at 4\mjup, where the interior equation of state changes 
from being dominated by electron degeneracy pressure to having a significant 
contribution by metallic hydrogen (Liebert's contribution in Boss et al. 2003).
In this paper we adopt the D-burning limit as the dividing line 
between BDs and planets, so that we are consistent with our previous work 
in this cluster. 

SMOs have also been found orbiting main-sequence stars. The mass distribution 
of companions to low-mass main-sequence stars within 3~AU shows that SMOs 
with masses between 70 and 20\mjup ~are rare, but their frequency increases 
 below about 10\mjup ~(Queloz, Santos and Mayor 2002). 
On the other hand, there are indications that BDs may be frequent at large separations from stars (Gizis et al. 2000; Potter et al. 2002).  
Theoretical work has shown that it is possible that planetary mass 
companions to stars form by gravitational collapse 
rather than by agglomeration of planetesimals (Boss 2001; Bate, Bonnell 
\& Bromm 2003).  
This scenario would seem more plausible if there were also free-floating 
planets that have directly formed from the collapse and fragmentation 
of molecular clouds, or that have been ejected by interactions with stars 
and other SMOs. 

The \sigori~ cluster is a prime location to search for SMOs because of 
its young age ($\sim$3~Myr), distance ($\sim$440~pc) and low reddening (Sherry 2003). 
A cluster of X-ray emitting young stars was discovered by Walter, Wolk \& Sherry (1998). 
A sequence of SMOs has been found extending to masses below the D-burning limit 
(Zapatero Osorio et al. 2000; B\'ejar et al.  2001). 
The coolest and faintest object so far detected 
in any young cluster or association is S\,Ori\,70 (Zapatero Osorio et al. 2002).
In this paper we report spectroscopic observations of S\,Ori\,70 that have more than a factor of 10  higher spectral resolution than our previous data. 
We use the higher resolution spectrum to improve previous estimates of the gravity and temperature by fitting with synthetic spectra. We confirm that S\,Ori\,70 has a low gravity, which strongly supports its membership in the 
\sigori~ cluster.

\section{Observations and data reduction}

On 2002 November 24th, we observed S\,Ori\,J053810.1-023626 
(hereafter S\,Ori\,70) with the Keck~II near-infrared 
spectrometer NIRSPEC (McLean et al. 1998). Meteorological conditions were slightly 
non-photometric (occasional light cirruses) 
and the seeing ranged from 0.6 arcsec to 0.9 arcsec, as measured in the NIRSPEC images. 

Total exposure times were 4200\,s and 600\,s for the $J$-band spectra
of S\,Ori\,70 (J=20.28, Zapatero Osorio et al. 2002) 
and 2MASS\,J0559191-140449, respectively. 
The latter object is a relatively bright T5 dwarf 
(J=13.82, Burgasser et al. 2002). 
The observing strategy employed for S\,Ori\,70 was as follows: two individual
integrations of 600\,s at three different positions along the entrance
slit separated by about 13\arcsec. An additional exposure of 600\,s
was obtained at one of the positions. Only two single integrations of
300\,s were collected for 2MASS\,J0559191-140449. In order to remove
telluric absorptions due to the terrestrial atmosphere, the
near-infrared featureless A0V-type stars HD\,15820 and HD\,63817 were
observed very close in airmass (typically within 0.05
airmasses). White light calibration images were taken at the beginning
and at the end of the observing night.

Raw data were reduced within {\sc iraf}\footnote{IRAF is distributed
by National Optical Astronomy Observatory, which is operated by the
Association of Universities for Research in Astronomy, Inc., under
contract with the National Science Foundation.} and following standard
techniques in the near-infrared. Nodded images were subtracted to
remove the sky background and dark current. The object spectra were
then optimally extracted using subroutines of the {\sc twodspec}
package. The extracted spectra were divided by their corresponding
normalized extracted flat-fields. Wavelength calibration
($rms$\,=\,3\,\AA) was performed using 33 OH sky emission lines. The
hydrogen P$\beta$ absorption line at 1.2818\,$\mu$m in the spectra of
the A0V-type stars was interpolated before the corresponding science
spectra were divided by them to cancel terrestrial features. To
complete the data reduction, we multiplied the spectra of our targets
by the black body spectrum for the temperature of 9480\,K, which
corresponds to the A0V class (Allen 2000). 

Our final NIRSPEC spectrum of  S\,Ori\,70 is shown in Figure~1. It has 
a resolving power of R=1500. 
We also show the spectrum of 2MASS\,J0559191-140449 obtained the same night. 
The shape of the spectra are different, indicating different physical 
conditions in the photosphere. Lucas et al. (2001) showed that triangular 
shapes in near-infrared spectra of L dwarfs could be explained by the low 
gravity of SMOs in the Trapezium. In the next section we analyze these 
conditions using theoretical tools. We find that the triangular 
shape of the J-band spectrum 
of  S\,Ori\,70 is caused by a surface gravity lower than that of 
2MASS\,J0559191-140449, 
consistent with having a larger radius as expected for a younger object. 

\section{Scientific analysis}

In our analysis we use synthetic spectra for T dwarfs that have been 
developed by Allard et al. (2001). They include the total condensation of 
dust grains below the photosphere (COND models) and the latest 
AMES list of H$_2$O lines (Partridge \& Schwenke 1997). 
These models reproduce well the blue 
near-infrared colors of T dwarfs in  the \sigori~ cluster 
(Mart\'\i n et al. 2001; Zapatero Osorio et al. 2002).  
In Figure~2, we show the effects of different surface gravities in the synthetic 
spectra calculated with COND atmospheric models. A triangular shape is expected 
to develop for lower gravities due to enhanced methane and steam absorption. 
The strength of the KI doublet at 1.25 microns is also greater for lower gravities. 

Our NIRSPEC spectrum of S\,Ori\,70 has been compared with a grid of theoretical spectra 
using a least square minimization technique. The best fit is shown in Figure~3. 
It was obtained for a COND
model with $T_{\rm eff}$\,=\,1100\,K and log\,$g$\,=\,3.5\,cm\,s$^{-2}$. 
We estimate that the internal 1$\sigma$ uncertainty of our fitting technique 
is  $^{+200}_{-100}$\,K and $\pm$\,0.5\,cm\,s$^{-2}$, respectively. 

Using the spectroscopically derived surface gravity and temperature, we can place 
S\,Ori\,70 in an evolutionary diagram that distance independent  (Figure 4). 
The object lies on the 3~Myr isochrone for a mass of 3 Jupiters using the 
evolutionary models of Burrows et al. (1997). A similar result was obtained from 
the models of Chabrier et al. (2000). 
Higher mass, older T dwarfs, have 
higher gravities, such as those inferred for the 
T dwarfs Gl\,229\,B and Gl\,570\,D, also shown in Figure~4. If S\,Ori\,70 were  
a brown dwarf in the line of sight to the \sigori~ cluster, it should have 
a higher gravity. The low gravity of this object implies a very young age that is 
fully consistent with the cluster's age (B\'ejar et al. 2001; Oliveira et al. 2002; 
Sherry 2003). 

For very young ages ($\le$1~Myr), the ages and masses derived from evolutionary 
models may depend strongly on the initial conditions assumed in the calculations 
(Baraffe et al. 2002). However, at an age of 3~Myr, the models used to 
obtain the age and mass of S\,Ori\,70 are thought to be safely independent of the 
initial conditions.

\section{Discussion}

Spectroscopic analysis of gravity sensitive features has been recognized 
as a powerful tool to estimate the masses of SMOs (Davidge \& Boeshaar 1991). 
As they contract with time, the gravity of SMOs changes by more than two orders of 
magnitude. Very young SMOs have low gravities (log~g$\sim$3.0), while old ones have 
high gravities (log~g$\sim$5.5). The strengths of the molecular absorption bands increase with decreasing 
surface gravity because of changes in the temperature structure of the atmosphere. 
At lower gravity a larger column abundance of molecules is required to 
compress the atmosphere to a given pressure.  
Spectral synthesis analysis of observed spectra have confirmed that lower gravity objects with  
late-M and L spectral types have stronger molecular bands (Allard et al. 2001; 
Basri et al. 2000; Gorlova et al. 2003; Leggett et al. 2001; 
Lucas et al. 2001; Mohanty et al. 2003; Schweitzer et al. 2001, 2002). 
As shown in Figure~2, gravity effects 
are also significant in the spectrum of T dwarfs. 

S\,Ori\,70 is currently the coolest object identified in any  cluster 
(Zapatero Osorio et al. 2002). Consequently, it deserves particular attention 
because it indicates that there could be a numerous population of planetary members 
of clusters.  The estimated mass of S\,Ori\,70 overlaps with the masses 
of planetary companions to stars, giving an impetus to the ideas that giant planets 
form by rapid gravitational processes that happen in molecular cloud cores, disks, and 
filaments (Boss 2001; Bate, Bonnell \& Bromm 2003). 

The angular distance of S\,Ori\,70 from the 
naked-eye O9.5V star $\sigma$ Ori (HR 1931) is 8'.65, corresponding to a projected 
separation of 182676~AU at 
352~pc. This star is actually a quintuple system where the most distant companion is 
a B2 star at 42". The total mass of these 5 stars is about 60~M$_\odot$. 
The maximum separation of binaries with O-type primaries is about 110000~AU 
(Abt 1986). Thus, the separation from the  star $\sigma$ Ori to the 
planetary object S\,Ori\,70 is of the same order of magnitude as the widest binaries 
with O-type primaries. 

Could S\,Ori\,70 be part of a planetary system surrounding the 
massive multiple stellar system?. Circumbinary disks are know to exist around 
low-mass stars (e.g., Potter et al. 2000). However, it is not known whether 
very large disk structures could exist around high-mass multiple stars.   
A dusty proto-planetary disk without a central star has recently been discovered 
at only 1200~AU from $\sigma$ Ori (van Loon \& Oliveira 2003), suggesting that planetary 
formation is still ongoing in this region. However, the typical separation 
between stars in the \sigori~ cluster is about 50000~AU. In fact, there are 3 brown 
dwarfs and one star within 65000~AU of S\,Ori\,70. We conclude that S\,Ori\,70 is likely a planetary  member in the cluster, but not particularly bound to any star or brown dwarf. 
Could S\,Ori\,70 have been ejected from an unstable multiple stellar or planetary system?. 
We do not have kinematic information so we cannot constrain this hypothesis. 
Nevertheless, we note that the escape velocity from the cluster 
(Vesc=0.65~km/s) is lower than the typical ejection 
velocities produced in the simulations of Bate et al. (2002). Most ejections occur very early 
in the evolution of the unstable systems (age$<$0.1~Myr), 
so it is unlikely that S\,Ori\,70 has been recently ejected.

The presence of planetary members in open clusters that are not bound to any particular 
star opens a new perspective to the diversity of planets in the universe. 
A matter of ongoing debate is the most suitable nomenclature for planetary objects 
that are not orbiting solar-type stars. 
We adhere to a wide use of the term "planet" that includes 
all objects that do not develop nuclear reactions in their 
cores, independently of where they may be found or how they formed. Qualifiers may be added to specify the environmental conditions, such as 
"pulsar planets", "solar-system planets", and "close-in planets" which have already been used in the literature. We propose to call planets such as S\,Ori\,70, which are located in a star cluster but 
do not orbit any particular star, "cluster planets". Our proposition differs from the  
definition of the IAU Working Group on Extrasolar Planets (Boss et al. 2003), which states that: 
"Free-floating objects in young star clusters with masses below the 
limiting mass for thermonuclear fusion of deuterium are not planets, but 
are sub-brown dwarfs (or whatever name is most appropriate)."

Cluster planets are probably numerous in the galaxy as suggested by the IMF presented in B\'ejar et al. (2001). The \sigori~ cluster is probably not bound because it does not have enough mass 
(Sherry 2003). 
Thus, cluster planets become "free-floating planets" after several million years. 
Free-floating planets are likely to populate the solar neighborhood. 
 
\acknowledgments

We thank France Allard for sending us synthetic spectra, and David Barrado y Navascu\'es, Victor S\'anchez B\'ejar and Mark Marley for comments that helped to improve the manuscript. 
MRZ was supported by the Ram\'on y Cajal postdoctoral program.  
This material is based upon work supported by the National Science
Foundation under Grant No. 0205862. 
Any opinions, findings, and conclusions or recommendations expressed
in this material are those of the author(s) and do not necessarily reflect
the views of the National Science Foundation. 
The authors wish to extend special thanks to those of Hawaiian ancestry on 
whose sacred mountain of Mauna Kea we are privileged to be guests. 
Without their generous hospitality, 
the Keck~II telescope observations presented therein would 
not have been possible.

\clearpage

\figcaption[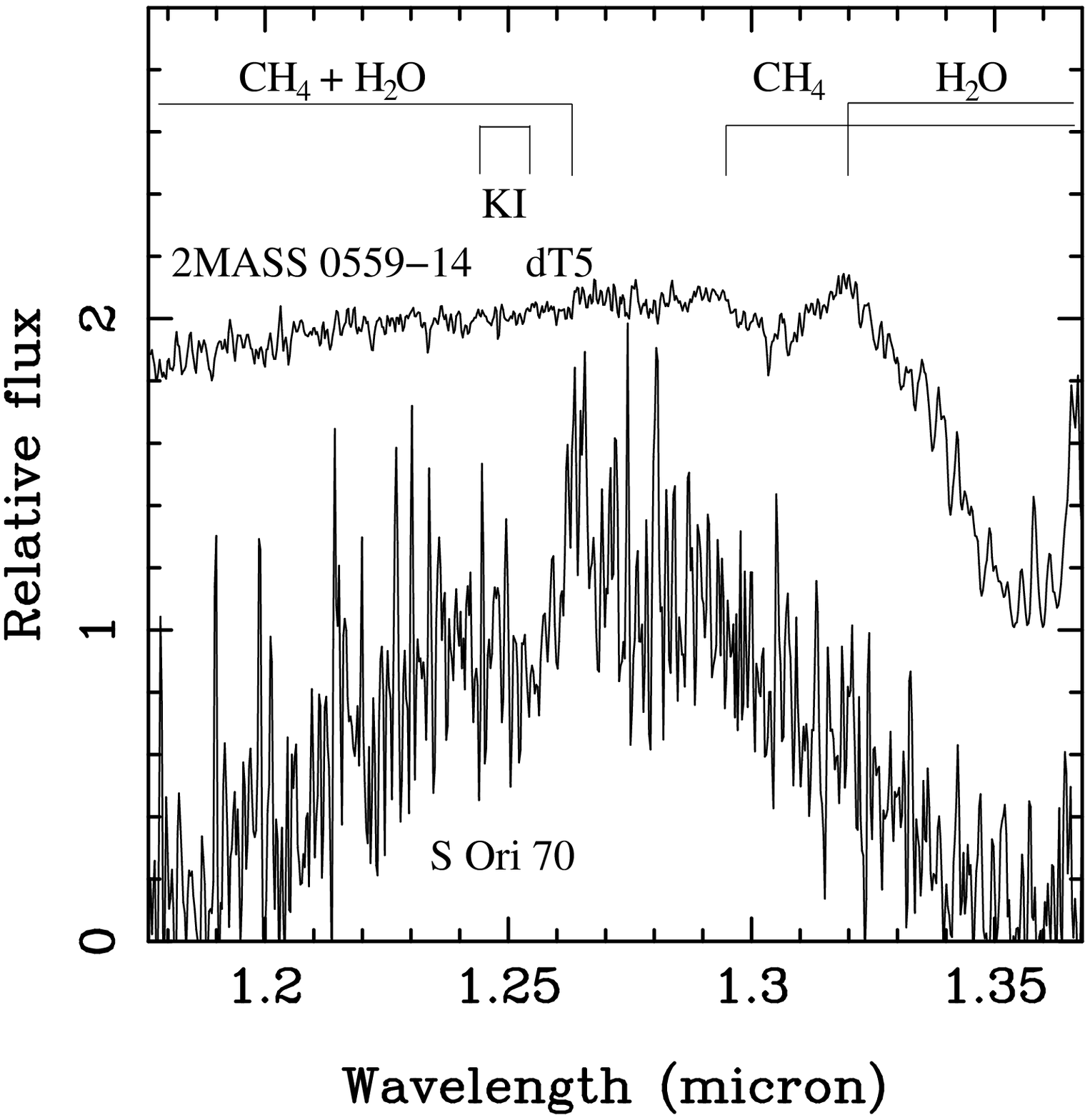]{\label{fig1} NIRSPEC $J$-band spectra of S\,Ori\,70 and 
the T5  field brown
dwarf 2MASS\,J0559191-140449 (Burgasser et al. 2000). 
Data have been normalized to unity at 1.235--1.24\,$\mu$m, and
have been shifted for clarity. Note the strong water vapor absorption
present in the spectrum of S\,Ori\,70, and the "flat" spectrum of the
field brown dwarf. Major molecular and atomic features are indicated.} 

\figcaption[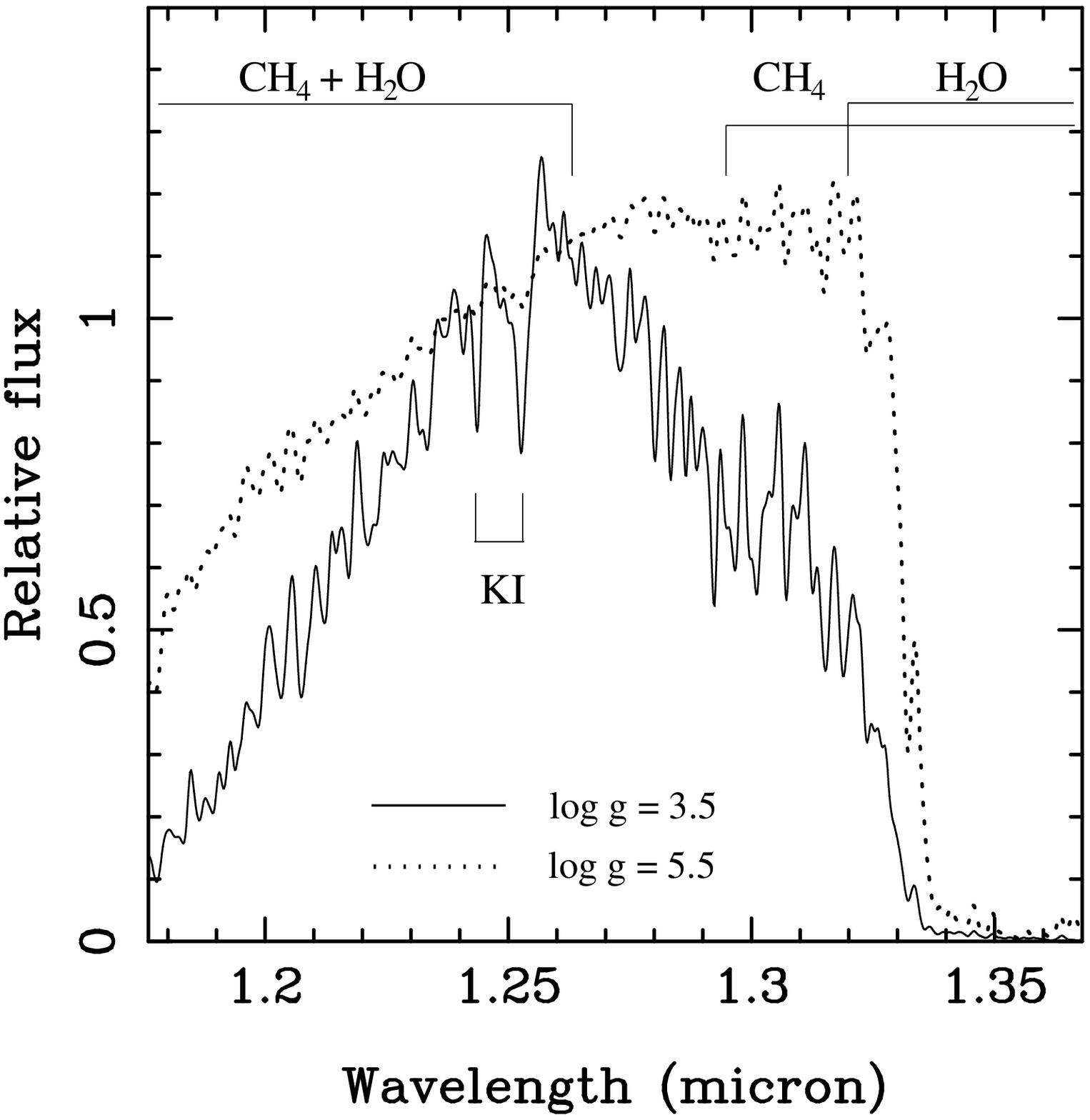]{\label{fig2} COND synthetic spectra computed for 
$T_{\rm eff}$\,=\,1000\,K and two
different gravities: log\,$g$\,=\,3.5 and 5.5 (cm\,s$^{-2}$). These
spectra were kindly provided by France Allard and are described in
Allard et al. (2001). At these wavelengths, the
low-gravity model shows stronger water vapor absorption and stronger
K\,{\sc i} lines than does the high-gravity model. Theoretical spectra
have been degraded to the resolution of NIRSPEC data (about 8\,\AA).} 

\figcaption[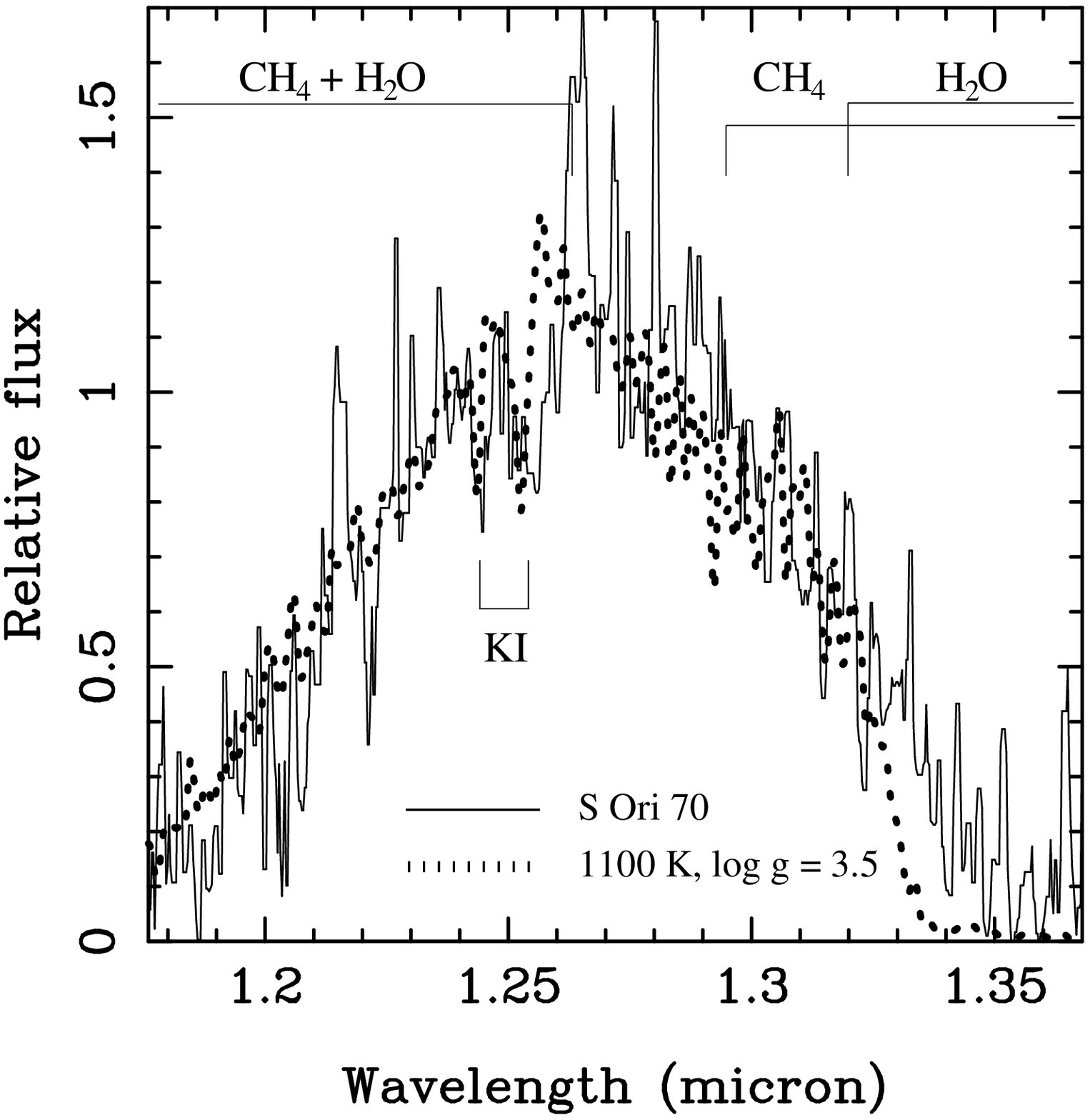]{\label{fig3} 
The NIRSPEC spectrum of 
S\,Ori\,70 (solid line) compared with the COND synthetic spectrum 
with $T_{\rm eff}$\,=\,1100\,K and
log\,$g$\,=\,3.5\,cm\,s$^{-2}$. As in Fig.~1, major
molecular and atomic features are indicated.} 

\figcaption[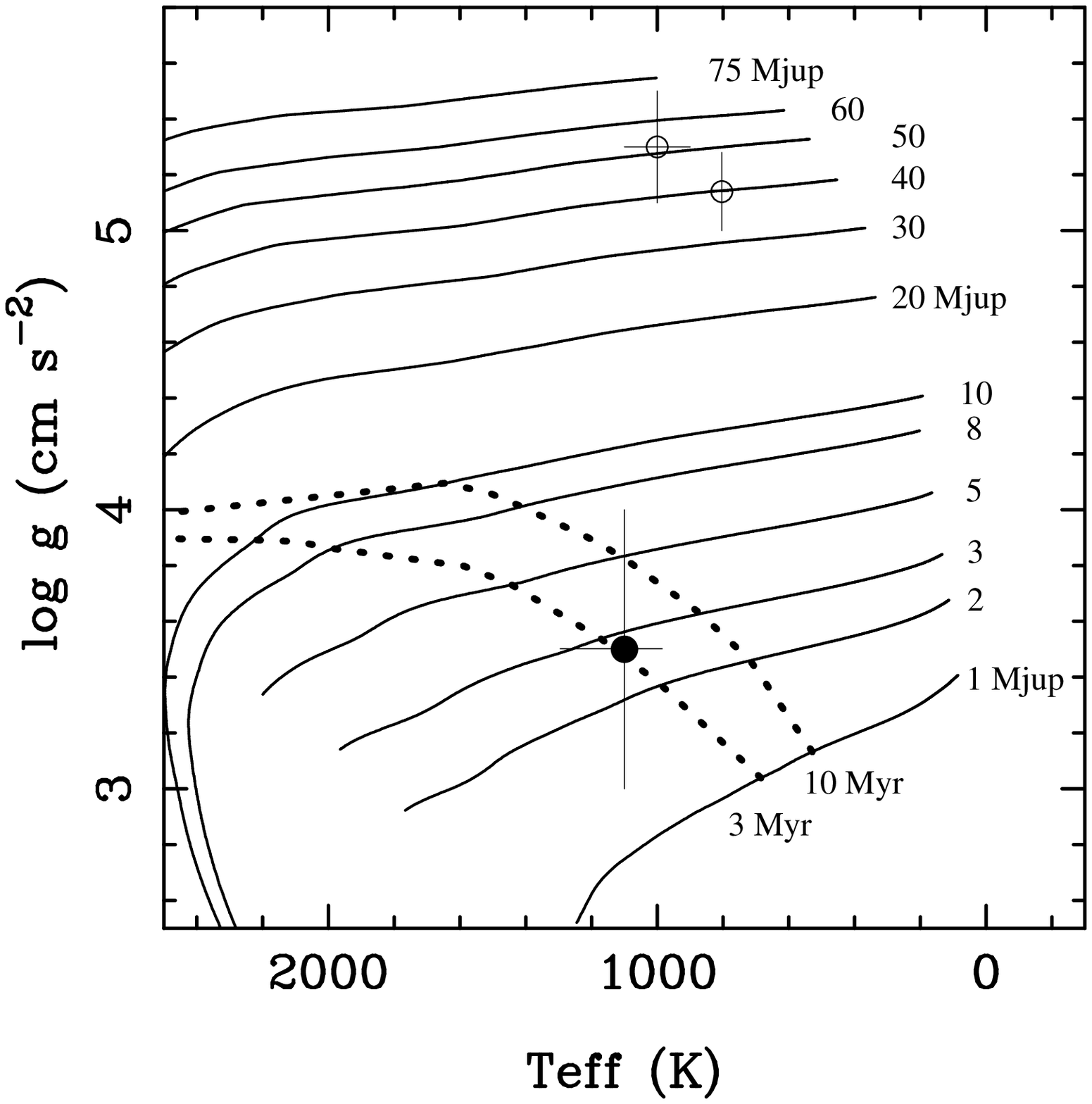]{\label{fig4} Location of S\,Ori\,70 (filled circle) 
in the gravity versus effective temperature 
diagram. Evolutionary models are taken from Burrows et al. (1997).  
From top to bottom, tracks for brown dwarf and planetary masses are
delineated with solid lines (masses are labelled in units of Jupiter).
The 3 and 10\,Myr isochrones are plotted with dotted lines. Also shown
with open circles are the locations of the T-class field brown dwarfs
Gl\,229\,B and Gl\,570\,D (Allard et al. 1996; Geballe et al. 2001).  
From the figure, we derive that S\,Ori\,70 is younger than 10\,Myr,
with a likely age of 3\,Myr, and has a planetary mass of
3\,$^{+5}_{-2}$ times the mass of Jupiter.}

\clearpage
\epsscale{0.7}
\plotone{f1.eps}
\clearpage
\epsscale{0.7}
\plotone{f2.eps}
\clearpage
\epsscale{1.1}
\plotone{f3.eps}
\clearpage
\epsscale{1.1}
\plotone{f4.eps}

\end{document}